\documentclass[aps,twocolumn,superscriptaddress]{revtex4-2} 
\usepackage{amssymb,amsmath,amsthm,amsfonts,amsbsy,mathrsfs}
\usepackage[english]{babel}
\usepackage{epsfig,amssymb,amsmath,amsthm,amsfonts,amsbsy,mathrsfs}
\usepackage{graphicx}
\usepackage{hyperref}
\usepackage{cleveref}
\usepackage{color}
\usepackage{bm}

\begin{document}

\title{Emergence of linear isotropic elasticity in amorphous and polycrystalline materials}

\author{Shivam Mahajan}
\affiliation{Division of Physics and Applied Physics, School of Physical and
Mathematical Sciences, Nanyang Technological University, Singapore}

\author{Joyjit Chattoraj}
\affiliation{Division of Physics and Applied Physics, School of Physical and
Mathematical Sciences, Nanyang Technological University, Singapore}
\affiliation{Institute of High Performance Computing, Agency for Science, Technology and Research, Singapore}

\author{Massimo Pica Ciamarra}
\email{massimo@ntu.edu.sg}
\affiliation{Division of Physics and Applied Physics, School of Physical and
Mathematical Sciences, Nanyang Technological University, Singapore}

\date{\today}
%
\begin{abstract}
We investigate the emergence of isotropic linear elasticity in amorphous and polycrystalline solids, via extensive numerical simulations.
We show that the elastic properties are correlated over a finite length scale $\xi_E$, so that central limit theorem dictates the emergence of continuum linear isotropic elasticity on increasing the specimen size.
The stiffness matrix of systems of finite size $L > \xi_E$ is obtained adding to that predicted by linear isotropic elasticity a random one of spectral norm $(L/\xi_E)^{-3/2}$, in three spatial dimensions.
We further demonstrate that the elastic length scale corresponds to that of structural correlations, which in polycrystals reflect the typical size of the grain boundaries and length scales characterizing correlations in the stress field.
We finally demonstrate that the elastic length scale affects the decay of the anisotropic long-ranged correlations of locally defined shear modulus and shear stress. 
\end{abstract}
\maketitle

\section{Introduction}
Linear isotropic elasticity (LIE) describes
the mechanical response of macroscopic molecular solids assuming matter to be continuous and rotationally invariant. 
These assumptions are not met at the microscopic scale.
Indeed, the elastic properties of small polycrystalline~\cite{Mullen1997} or amorphous~\cite{Wittmer2002} samples exhibit large sample-to-sample fluctuations.
Similar size fluctuations characterize the elastic response in the plastic regime, where they have been extensively investigated (see, e.g.~\cite{Sethna2017}). 
The elastic response fluctuations vanish as the linear size of a sample increases and LIE becomes more accurate.
Accordingly, LIE's validity depends on the ratio between the linear system size, $L$, and a microscopic elastic length scale, $\xi_E$. 
What sets this length scale? And how does the validity of LIE depends on $L/\xi_E$?
These questions have been separately addressed in amorphous or polycrystalline materials.

For amorphous solids, extensive simulations have investigated the convergence of the elastic response to linear isotropic elasticity in model Lennard-Jones like systems. 
Tanguy~\cite{Tanguy2002} et al. found the stress anisotropy to decrease exponentially with the system size with a decay length of the order of $65$ particle diameters, which is a possible estimation of $\xi_E$.
This length scale has been associated with the correlation length of the non-affine particle displacements induced by external deformations, which is also, typically, of the order of several diameters~\cite{Tanguy2002, Wittmer2002,Leonforte2005}.
Subsequent work~\cite{Tsamados2009} showed that the eigenvalues of the stiffness tensor evaluated over a coarse-graining length scale $w$ converge to their asymptotic limit as a power-law 
not complying with the central limit theorem expectation, and possibly dependent on the degree of structural order~\cite{Cakir2016}. 
We note, however, that these results may depend on the chosen definition of coarse-grained elastic quantities~\cite{Mizuno2013}. 

For polycrystals, the question of how the validity of LIE depends on $\xi_E/L$ has not been addressed. Previous works, indeed, mostly investigated how the elastic properties relate to those of the single grains in the limit $L/\xi_E \gg 1$, e.g. through the Voight~\cite{Voigt1889} or Reuss~\cite{Reuss1929} averages or more refined approaches~\cite{Mavko2009, Avellaneda1996}.
In polycrystals, the length scale $\xi_E$ is heuristically identified with the typical grain size~\cite{Chaikin2010}, despite concerns on the connection between structural and elastic length scales~\cite{Nagel}. 

In this paper, we investigate the emergence of LIE in materials with different degree of structural disorder, from amorphous to polycrystalline, produced via large-scale three-dimensional numerical simulations of the cooling process of liquid samples, at different cooling rates (Sec.~\ref{sec:model}).
We demonstrate in Sec.~\ref{sec:lie} that deviations from LIE scales with the linear size $L$ of the system as $(L/\xi_E)^{-3/2}$, where $\xi_E$ is an elastic correlation length.
This result implies that finite-size effects act as a random perturbation to the stiffness matrix, as we discuss in Sec.~\ref{sec:perturbation}.
We further show in Sec.~\ref{sec:structural} that the correlation length $\xi_E$, that grows as the cooling rate decreases, (i) corresponds to a structural correlation length $\xi_S$ which for polycrystalline materials coincides with the grain size and (ii) controls the size dependence of the pressure and anisotropy of the stress tensor.
Finally, in Sec.\ref{sec:local} we study the correlation of locally defined stress and compliance tensors.
We show that these tensors are characterized by long-ranged anisotropic correlations, confirming previous findings
~\cite{Lemaitre2014,Wu2015a,Lemaitre2015,Lemaitre2018}, and show that the decay of these correlations are governed by the elastic length scale $\xi_E$.

\section{Numerical model and protocols \label{sec:model}}
\begin{figure}[t!]
 \centering
 \includegraphics[angle=0,width=0.48\textwidth]{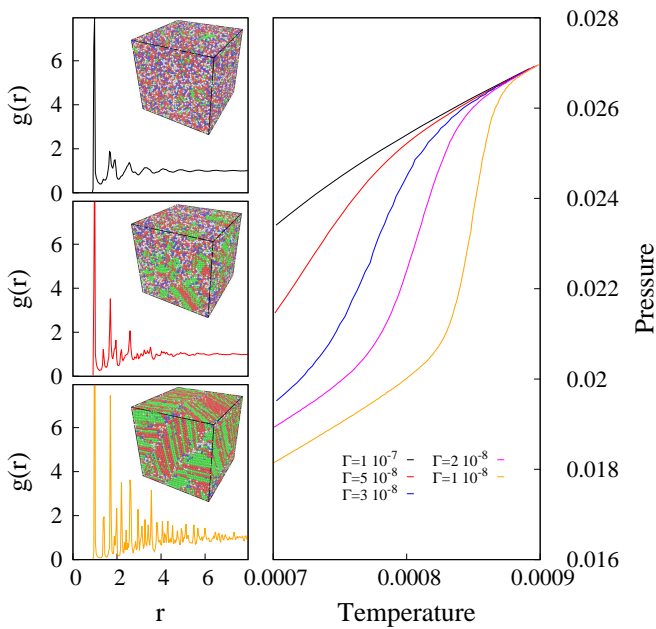}
 \caption{
 We illustrate in the right panel the dependence of the pressure on the temperature, for different cooling rates. Energy minimization of the $T=7\times10^{-4}$ configuration bring the system in solid states with different degree of disorder.
 The left panels illustrate example radial distribution functions and snapshots of these solids, for $N=131072$.  
 The color of a particle identifies its local crystal structure~\cite{Ackland2006,Stukowski2010}, fcc (green), hpc (red), bcc (blue), icosahedral (yellow), none (gray).
  In this work, we investigate the elastic properties of these solids, in the linear response regime.
  \label{fig:cooling}
  }
\end{figure}

We perform large-scale numerical simulations of monodisperse spherical particles of diameter $\sigma$ interacting via the Hertzian potential, $v(r)= \frac{2}{5}\epsilon(r-\sigma)^{5/2}$ for $r < \sigma$, $v(r) = 0$ otherwise. 
We fix the volume fraction to $\phi=0.74$, a value at which the ground state is an fcc crystal~\cite{Pamies2009}, and prepare solid samples by quenching equilibrated liquid configurations to low temperature, using periodic boundary conditions.
We mimic quenches to temperatures well below the melting one, $T_m$, by first cooling the system to $T_l \simeq 0.8 T_m$ at rate $\Gamma$, and then minimizing the energy via the conjugate-gradient algorithm. 
The cooling rate affects the ordering properties of the resulting configuration, which is amorphous at large $\Gamma$, and polycrystalline at small $\Gamma$, as apparent from Fig.~\ref{fig:cooling}.
For each cooling rate $\Gamma$ and number of particles $N$, in the range $500$ to $1$ million, we prepare $50$ independent samples.
All data reported in the following are averaged over these samples.

\section{Emergence of Linear isotropic elasticity~\label{sec:lie}}
According to LIE, in three dimensions, the stress-strain relation $\hat {\mathbf \sigma} = \hat {\bf C} \hat {\bf \epsilon}$ is,
\begin{equation}
\begin{pmatrix}
\sigma_{1} \\
\sigma_{2} \\
\sigma_{3} \\
\sigma_{4} \\ 
\sigma_{5} \\
\sigma_{6} 
\end{pmatrix}
=
\begin{pmatrix}
\lambda + 2\mu & \lambda & \lambda & 0 & 0 & 0\\
\lambda & \lambda + 2\lambda & \lambda & 0 & 0 & 0\\
\lambda & \lambda & \lambda + 2\mu & 0 & 0 & 0\\
0 & 0 & 0 & 2\mu & 0 & 0\\
0 & 0 & 0 & 0 & 2\mu & 0\\
0 & 0 & 0 & 0 & 0 & 2\mu
\end{pmatrix}
\begin{pmatrix}
\epsilon_{1} \\
\epsilon_{2} \\
\epsilon_{3} \\
\epsilon_{4} \\ 
\epsilon_{5} \\
\epsilon_{6} 
\end{pmatrix}
\label{eq:lietensor}
\end{equation}
Here, the suffix 1-6 indicates $xx,yy,zz,xy,xz,yz$ so that, e.g., $c_{14}$ stands for $c_{xxxy}$. 
The parameters $\lambda = \frac{\nu E}{(1+\nu)(1-2\nu)}$ and $\mu = G = \frac{ E}{2(1+\nu)}$ are the Lam\'e constants, and $E$, $G$ and $\nu$ are the Young modulus, shear modulus and Poisson ratio, respectively. 
If LIE holds, therefore, the six invariants of the stress tensor $\hat {\bf C}$ are  $2\mu$, with multiplicity five, and $3\lambda +2\mu$, with single multiplicity.
However, in finite systems rotational invariance is broken and hence $\hat {\bf C}$ is a symmetric matrix with entries depending on the reference frame.
A frame-independent evaluation of the LIE's validity~\cite{Tsamados2009} is thus obtained comparing the invariants of $\hat {\bf C}$ with those predicted by LIE.

To evaluate the stiffness matrix, we impose to each configuration a strain deformation followed by energy minimization. 
We perform this operation for the six deformation modes, $d(\epsilon_{\alpha\beta})$.
In the linear response regime, which we have checked to occur for strains $d(\epsilon_{\alpha\beta}) \lesssim 10^{-7}$, this allows evaluating the stiffness matrix $c_{\alpha\beta\gamma\delta}$ from the changes in stress tensor, $d(\sigma_{\alpha\beta})$,
\begin{equation}
c_{\alpha\beta\gamma\delta}(N) = \frac{d(\sigma_{\alpha\beta})}{d (\epsilon_{\gamma\delta})}
  \label{eq:ds}
\end{equation}
The subsequent diagonalization of the stiffness matrix yields six eigenvalues, we indicate with $c_1 \leq \ldots \leq c_5 \leq b$. 

\begin{figure}[t!]
 \centering
 \includegraphics[angle=0,width=0.48\textwidth]{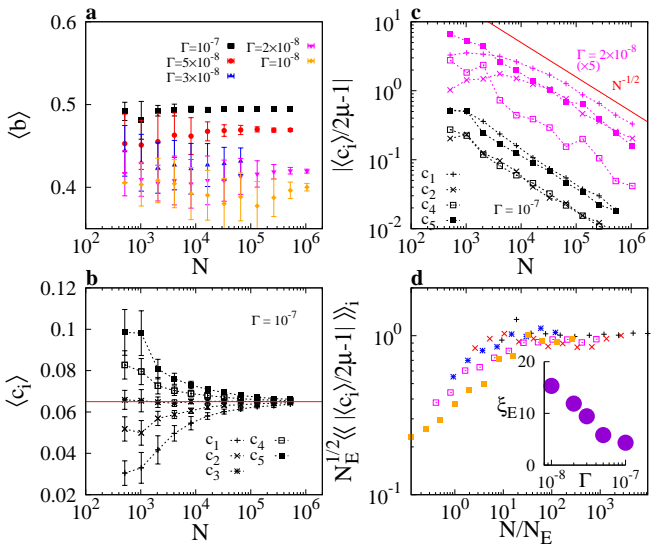}
 \caption{
  System size dependence (a) of the largest eigenvalue of the stiffness matrix for different cooling rates, and (b) of the other $5$ eigenvalues $c_1,\ldots,c_5$ for $\Gamma = 10^{-7}$. 
  Panel (c) illustrates the approach of the average of each $\langle c_i \rangle$ to the common asymptotic value $2\mu$ on increasing the system size, for $\Gamma = 10^{-7}$ and $\Gamma = 10^{-8}$. 
  Data are averaged over $50$ realizations for each system size and cooling rate.
  Panel (d) illustrates that the average 
  of the 5 eigenvalues $\langle c_i \rangle$ approach its asymptotic limit as $\left(N_E/N\right)^{1/2}$.
  This allows defining an elastic length scale $\xi_E = N_E^{1/3}$ which grows as $\Gamma$ decreases, as illustrated in the inset.
  \label{fig:eigs}
  }
\end{figure}

We observe the sample average of the largest eigenvalue $\langle b \rangle$ to become asymptotically size independent, $\langle b \rangle-(3\lambda+2\mu)\propto N^{-k_b}$ with $k_b \simeq 0$, as illustrated in Fig.~\ref{fig:eigs}a. 
$\langle b \rangle$ decreases with $\Gamma$, a finding explained considering that, at constant volume, ordered systems have a smaller pressure, as in Fig.~\ref{fig:cooling}. 
When the effect of pressure is filtered out investigating $\langle b \rangle/\langle P \rangle$, ordered systems result stiffer than disordered ones.
At each $\Gamma$, the five eigenvalues $\langle c_i\rangle$ approach a common limiting value $2\mu$ as the system size increases. We find, in particular, that $c_1$ and $c_2$ approach the asymptotic value from below, $c_4$ and $c_5$ from above, while $c_3 \simeq 2\mu$ regardless of the system size. 
As an example, we illustrate the size dependence of the eigenvalues in Fig.~\ref{fig:eigs}b, for $\Gamma = 10^{-7}$.
The eigenvalues approach their common asymptotic limit as 
\begin{equation}
|\langle c_i\rangle-2\mu| = 2\mu \left(\frac{N}{{N}_E}\right)^{-k_c},
\label{eq:pl}
\end{equation}
with $k_c = 1/2$ and ${N}_E$ slightly dependent on the considered eigenvalue, as illustrated for $\Gamma = 10^{-7}$ and $\Gamma = 2\times 10^{-8}$ (data scaled by a factor $5$) in panel c.

For each cooling rate, we also compute $\langle\hspace{-0.05cm}\langle |\langle c_i\rangle-2\mu| \rangle\hspace{-0.05cm}\rangle$, where $\langle\cdot \rangle$ denotes an average over different realizations, and $\langle\hspace{-0.05cm}\langle\cdot \rangle\hspace{-0.05cm}\rangle$ averages over the different eigenvalues. 
Fig.~\ref{fig:eigs}d shows that this quantity scales as $N^{-1/2}$ for $N > N_E$, with $N_E$ increasing as the cooling rate decreases, as in Fig.~\ref{fig:eigs}d.
We remark that $N_E$ can be identified with the disorder parameter introduced by fluctuating elasticity theory~\cite{SchirmacherPRL, SchirmacherEPL2006, Marruzzo2013a,Shivam2}.
Furthermore, we notice that these findings are in line with previous results on the dependence of the sample-to-sample fluctuations of the elastic constants on the systems size~\cite{Mizuno2016a,Mizuno2016b,kapteijns2020elastic}.

However, these results represent a significant departure from previous findings~\cite{Tsamados2009} on the dependence of the stiffness matrix's eigenvalues on a coarse-grained length scale, $w$. 
Indeed, this previous work in two-spatial dimensions, found  the largest eigenvalue to approach its asymptotic limit as $w^{-2}$, and the other two as $w^{-0.87}$. 
By associating the exponents to volume and surface effects~\cite{Tsamados2009}, the scalings should be $w^{-3}$ and $w^{-2}$, in three spatial dimensions, corresponding to $k_b = 1.5$ and $k_c=1$, in marked contrast with our findings, $0$ and $0.5$, respectively. 

\section{Size effects as  perturbations~\label{sec:perturbation}}
\begin{figure}[t!]
 \centering
 \includegraphics[angle=0,width=0.48\textwidth]{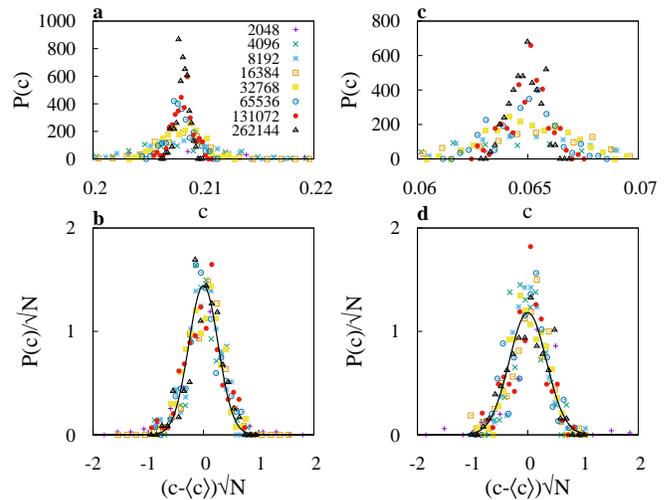}
 \caption{
 Panels (a) and (c) illustrate the distribution of the tensor elements  $c_{11}, c_{22}, c_{33}$ and $c_{44}, c_{55}, c_{66}$, respectively, for different system sizes, at cooling rate $10^{-7}$. 
 These distributions are respectively collapsed in (b) and (d).
 The full lines are Gaussian fits to the $N=65536$ data.
  \label{fig:pscaling}
  }
\end{figure}

We rationalize our findings considering that the stress change resulting from an applied deformation is 
\begin{equation}
d (\sigma_{\alpha\beta}) =
  d (\epsilon_{\gamma\delta}) c_{\alpha\beta\gamma\delta}(N) = \frac{\rho}{N} \sum_i d (r_\alpha f_\beta)_i,
  \label{eq:dc}
\end{equation}
where $r_\alpha$ and $f_\beta$ are the $\alpha$ and $\beta$ components of the distance and the interaction force of the particles involved in bond $i$, respectively, where a bond correspond to an interparticle interaction.
Since the strain is given, each matrix element $c_{\alpha\beta\gamma\delta}(N)$ is the average of $\propto N$ numbers.
If the contributions $d (r_\alpha f_\beta)_i$ are asymptotically uncorrelated, then by central limit theorem each matrix element is Gaussian distributed with average $c_{\alpha\beta\gamma\delta}$, its expected value in the thermodynamic limit, and variance scaling as $N^{-1/2}\propto L^{-d/2}$, in $d$ spatial dimensions.
Indeed, we observe in Fig.~\ref{fig:pscaling} that the distributions of the matrix elements collapse on a Gaussian curve when appropriately scaled. 
We remark that these collapses only occur asymptotically, $N > N_E$, implying the existence of short-ranged spatial correlations between the contributions of the different contacts to the stiffness matrix.

These findings imply that, for $N > N_E$, the stiffness matrix of a given realization is
\begin{equation}
{\hat {\bf C}}(N) = {\hat {\bf C}}(\infty) + \frac{1}{\sqrt{N}} {\hat {\bf R}},
\label{eq:lieN}
\end{equation}
where ${\hat {\bf C}}(\infty)$ is as in Eq.~\ref{eq:lietensor}, and ${\hat {\bf R}}$ is a Hermitian random matrix, with some given probability distribution and norm. 
Finite-size effects, therefore, are equivalent to a random perturbation of the asymptotic stiffness matrix.
Matrix perturbation theory~\cite{Stewart1990} then implies that each eigenvalue of ${\hat {\bf C}}(N)$ differs from its asymptotic limit by a constant proportional to the spectral norm of the perturbation, $N^{-1/2}$, as we have observed.

This theoretical interpretation allows rationalizing the results of Fig.~\ref{fig:eigs}, where we investigate how the averages of the {\it sorted} eigenvalues of the perturbed matrix approach their asymptotic values.
In a given realization, eigenvalue $b$, which is the largest, equals $b(N) = b_\infty + x$, where $x$ is random number of zero mean and standard deviation $\propto N$. 
The average over different configurations is therefore $\langle b \rangle(N) =b_\infty$: the average has no size dependence, i.e. $k_b = 0$, consistent with our observation in panel a.
The other five eigenvalues coincide in the thermodynamic limit.
At any finite $N$, noise splits their values, and the eigenvalues equal $c_i = c_\infty + x_i$, $i = 1,5$, where $x_i$ are random variables of zero mean and standard deviation $\propto N^{-1/2}$. 
Since we sort the eigenvalues, $x_i < x_{i+1}$, we have $|\langle c_i \rangle - c_\infty| \propto N^{-k_c}$, with $k_c = 1/2$, for $i \neq 3$. 
Conversely, for $i = 3$ we predict $\langle c_i \rangle = c_\infty$. 
All of these predictions are in agreement with our findings in Fig.~\ref{fig:eigs}.

In two dimensions, where the stiffness matrix has three eigenvalues, we predict $k_b = 0$, and $k_c=1/2$ for $i=1,2$.
This prediction for $k_c$ is in rough agreement with previous results~\cite{Tsamados2009}, which have reported $2k_c = 0.87$. 

\section{Mechanical and structural length scales~\label{sec:structural}}
\subsection{Elastic length scale}
The above results imply that the emergence of LIE is characterized by a typical size $N_E$, to which we associate a length scale $\xi_E := N_E^{1/d}$.
For $N>N_E$, the probability distributions of different matrix elements is Gaussian, and  Eq.~\ref{eq:pl} holds. 
This length scale measures the spatial correlation of different contacts' contributions to the stiffness matrix, $d(r_\alpha f_\beta)_i$.

Here, we extract this length scale via the linear regression fits shown in  Fig.~\ref{fig:eigs}d. 
The length scale $\xi_E$ grows as the cooling rate $\Gamma$ decreases and the system becomes more ordered.
It varies from $\xi_E \simeq 4\sigma$ at $\Gamma = 10^{-7}$ to $\xi_E \simeq 15\sigma$ at $\Gamma = 10^{-8}$. 

\subsection{Structural length scale}
In polycrystalline materials agglomerate of randomly oriented grains, $\xi_E$ is expected to correspond to the typical grain size. 
In amorphous materials, $\xi_E$ may reflect a structural length scale of difficult definition. 
Since the correlation between mechanical and geometrical properties of solids is debated~\cite{Nagel}, it is also possible that $\xi_E$ do not have a structural interpretation.

Here, we investigate the correlation between the elastic and the structural properties of our systems by associating to each particle its Steinhardt~\cite{Steinhardt1983} order parameters, $q_{lm}(i)=\frac{1}{N_b(i)}\sum Y_{lm}(\hat {\bf r_{ij}})$, where the sum runs over all $N_b$ neighbors of particle $i$, and $Y_{lm}(\hat {\bf r_{ij}}) = Y_{lm}(\theta_{ij},\psi_{ij})$ are the spherical harmonics.
We identify the neighbours through a Voronoi tessellation.
The scalar product 
$s_{ij} = \sum_{m=-6}^6 q_{6m}(i) q_{6m}^*(j)$ measures the correlation between the structures surrounding particles $i$ and $j$~\cite{Lechner2008}.
Hence, the decay of correlation function
\begin{equation}
S(r) = \frac{\sum_i \sum_j s_{ij} \delta(r-r_{ij})}{\sum_i \sum_j \delta(r-r_{ij})}
\label{eq:S}
\end{equation}
allows estimating a structural correlation length.

\begin{figure}[t!]
 \centering
 \includegraphics[angle=0,width=0.48\textwidth]{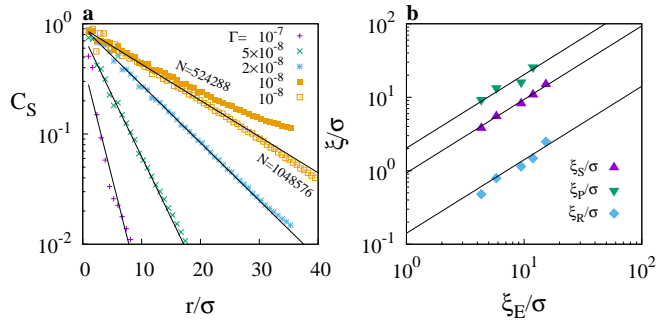}
 \caption{
  Structural correlation function (a), Eq.~\ref{eq:S}, for different cooling rates and $N = 524288$; for $\Gamma = 10^{-8}$, we also consider a larger $N$ value, as indicated.  The exponential decay of the correlation functions defines a structural length scale, $\xi_S$. Panel b shows that $\xi_S$, and the length scales $\xi_P$ and $\xi_R$ respectively associated to the pressure and to the stress anisotropy are proportional to $\xi_E$.
  \label{fig:length}
  }
\end{figure}

We find the correlation function $S(r)$ to decay exponentially, $S(r) = \exp(-r/\xi_S)$, with a characteristic structural length scale $\xi_S$ depending on the cooling rate, as shown in Fig.~\ref{fig:length}a.
Deviations from the exponential behaviour results from finite-size effects. 
The elastic length scale $\xi_E$ and the structural length scale $\xi_S$ turns out to be proportional, as illustrated in Fig.~\ref{fig:length}b. 
This result demonstrates a close connection between structural and elastic properties, equally valid in our polycrystalline and disordered systems.

\subsection{Stress length scale}
\begin{figure}[!!t]
 \centering
 \includegraphics[angle=0,width=0.48\textwidth]{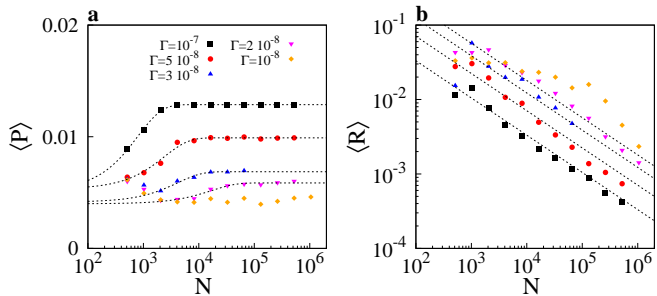}
 \caption{
  The average pressure approaches a liming value as the system size increases (a). The size dependence is well described by an exponential law, $P = P_0 + \Delta P e^{-N/N_P}$ (lines). The averaged stress anisotropy parameters exponentially scales as $(N/N_R)^{-1/2}$ (b).
  \label{fig:stress}
  }
\end{figure}

Microscopically, $\xi_E$ is the correlation length between the contribution of different interparticle contacts to the stiffness matrix, $d (r_\alpha f_\beta)_i$, Eq.~\ref{eq:dc}.
One may, therefore, wonder if the contributions $(r_\alpha f_\beta)_i$ of the contacts to the stress are similarly correlated.
We investigate this issue focussing on the dependence of average pressure $\langle P \rangle$ on the system size. Fig.~\ref{fig:stress}a illustrates that the average pressure exponentially approaches its asymptotic value as $N$ increases.
This allows defining a typical size $N_P$, and hence a typical pressure length scale $\xi_P := N_P^{1/d}$, we show to be proportional to $\xi_E$ in Fig.~\ref{fig:length}b.
We remark here that, for $\Gamma = 10^{-8}$, the pressure dependence on $N$ is too weak to allow for a reliable estimation of $\xi_P$.

Furthermore, we evaluate the degree of anisotropy of the stress tensor through the parameter $R = \sqrt{2 J_2}/P$, where $J_2$ is the second invariant of the deviatoric stress. 
Regardless of the cooling rate, $\langle R \rangle$ asymptotically scales as $(N/N_R)^{-1/2}$, as we illustrate in Fig.~\ref{fig:stress}b.
The corresponding length scale $\xi_R := N_R^{1/d}$ is also proportional to $\xi_E$, as we illustrate in Fig.~\ref{fig:length}b.

\section{Local elasticity~\label{sec:local}}
\begin{figure}[b!]
 \centering
 \includegraphics[angle=0,width=0.48\textwidth]{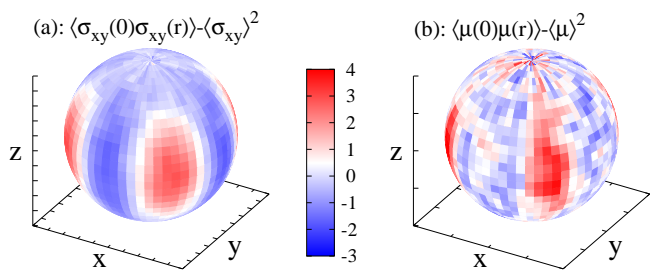}
 \caption{
  (a) Spherical map of the correlation function of the particle-level stress, $\sigma_{xy}({\bf 0})\sigma_{xy}({\bf r})$, for $|{\bf r}| \simeq 1.5$. 
  Data are normalizing using their standard deviation.
  (b) As in (a), but for the correlation function of the component $c_{44}$ of the particle-level stiffness matrix.
  \label{fig:spherical}
  }
\end{figure}
We now consider the possibility of extracting the elastic length scale via the direct study of the local elastic properties, rather than resorting to finite-size investigations.
To this end, we associate to each particle stress and elasticity tensors.
We define the stress tensor of particle $i$ as $\sigma_{\alpha\beta}^{(i)} = \frac{\rho}{2}\sum_j^{(i)} (r_\alpha f_\beta)_j$ where the sum is over all interaction forces involving particle $i$. 
We define a particle-level stiffness tensor $c^{(i)}_{\alpha\beta\gamma\delta}$ as 
$d\sigma_{\alpha\beta}^{(i)}/d(\epsilon_{\gamma\delta})$.
These two definitions, and in particular the adoption of a uniform strain, ensure that the macroscopic stress and stiffness tensors emerge as the average of the local ones.

We illustrate in Fig.~\ref{fig:spherical} spherical maps of the correlations functions of the local shear stress, $\langle \sigma_{xy}(r)\sigma_{xy}(0)\rangle - \langle \sigma_{xy}\rangle^2$ (a), and of $c_{44}$, we will refer to as the local shear modulus $\mu$, $\langle \mu(r)\mu(0)\rangle-\langle \mu \rangle^2$ (b) at $r \simeq 1.5$, for a $N=131072$ particle system in a disordered state, as obtained using the fastest of our cooling rates. 
The standard deviation of the correlations at the considered radial distance is used as a normalization factor.
In accordance with previous results ~\cite{Lemaitre2014,Wu2015a,Lemaitre2015,Lemaitre2018} this investigation evidences Eshelby-like quadrupolar anisotropic correlations both in the stress and in the local shear modulus.
\begin{figure}[t!]
 \centering
 \includegraphics[angle=0,width=0.48\textwidth]{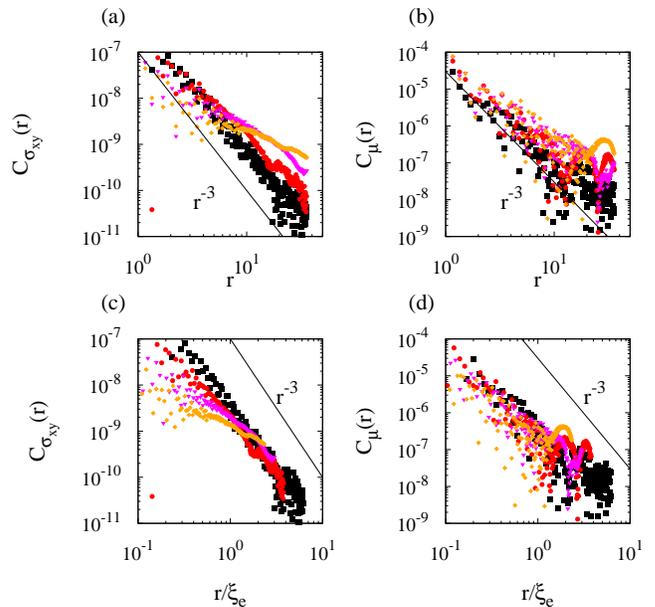}
 \caption{
  Correlation function of the particle defined $\sigma_{xy}$ for different cooling rates, plotted as a function of $r$ (a) and of $r/\xi_e$ (c).
  Analogous results for the correlation function of the particle defined $c_4$ are in panels (b) and (d), respectively.
  The correlation functions are averaged taking into consideration the quadrupolar symmetry of the fields. Symbols are as in Fig.~\ref{fig:stress}.
  \label{fig:corr3d}
  }
\end{figure}

We investigate the radial dependence of the observed stress correlations through ~\cite{Tong2020} an angle averaged correlation function,
$C_{\sigma_{xy}}(r) = -\frac{1}{2\pi}\int_0^\pi d\phi \int_0^{2\pi}d\theta [\langle \sigma_{xy}(r)\sigma_{xy}(0)\rangle - \langle \sigma_{xy}\rangle^2]$.
The correlation function $C_{\mu}(r)$ of the local shear modulus is similarly defined.
Fig.~\ref{fig:corr3d}a illustrates that 
$C_{\sigma_{xy}}(r) \propto r^{-3}$, after a transient, regardless of the cooling rate.
A similar result holds for the local shear modulus's correlation function, as illustrated in panel (b).
These results confirm the existence of long-ranged anisotropic correlations~\cite{Lemaitre2014, Wu2015a, Lemaitre2015, Lemaitre2018} in the stress and stiffness fields of amorphous materials.

When the correlation functions are plotted versus the radial distance scaled by the elastic length scale, as in  Figs.~\ref{fig:corr3d}(c) and (d), data for different cooling rates collapse in the asymptotic regime, within our numerical uncertainty.
This result indicates that the correlation functions asymptotically decays as $(r/\xi_e)^{-3}$, demonstrating how the elastic length scale can be evaluated from the analysis of locally defined elastic quantities.

We finally remark that self-averaging, the scaling of the fluctuations of the elastic properties with $N^{-1/2}$ (Fig.~\ref{fig:eigs}), holds as these long-ranged correlations are anisotropic in space. 
Positive and negative contributions cancels when evaluating the fluctuations via a volume integral of the correlation function.

\section{Conclusions} 
Our results establish that the emergence of isotropic linear elasticity is governed by central limit theorem, which sets in systems larger than a typical elastic length scale.
The existence of a finite correlation length in the elastic properties is in general agreement, e.g., with the assumptions of fluctuating elasticity theory~\cite{SchirmacherPRL, SchirmacherEPL2006, Marruzzo2013a}, as well as with the size dependence of the shear modulus reported in previous works~\cite{Mizuno2016a,Mizuno2016b,Lerner2019,kapteijns2020elastic}. 
The degree of disorder does not qualitatively affects this scenario, but influences the value of the elastic length scales.
Specifically, the elastic length scale grows with the degree of ordering and can be identified with the size of the grain boundaries, in polycrystalline materials.
We have further demonstrated that the elastic length scale, which we have derived via a finite size scaling investigation, can alternatively be measured via the study of the spatial correlation of locally defined elastic properties.

Either the finite-size scaling and real space investigations indicate that the correlation of the elastic properties reflect those of the frozen in stress. 
This is a result of practical significance, as correlations in the stress are easier to investigate than correlations in the local elastic constants.

We suspect that the structural correlation function we have introduced may be inappropriate in the presence of polydispersity or non-radially symmetric interaction potentials. 
In these cases where it is not apparent what structural correlation function relates to the elastic response. 
Possibly, in these cases structural correlations could be more meaningfully indirectly evaluated studying the correlation of the elastic properties.
This appears a promising direction to extract a static length scale in disordered materials whose relevance to, e.g. the glass transition problem~\cite{Karmakar} or plastic response~\cite{Sethna2017}, needs to be systematically explored.

In this regard, it is interesting to contrast our results with size-scaling studies of the  fluctuations of the shear modulus in systems whose crystallization is severely inhibited.
These studies considered systems first thermalised at a parent temperature $T_p$, and then brought to an energy minimal configuration.
The parent temperature, therefore, qualitatively plays the role of our cooling rate.
While we have observed that the elastic length scale grows as a system is better annealed, being correlated to the size of the grain boundaries, these previous studies have conversely found it to decrease~\cite{Rainone2020,Shivam2}. 
Recent results~\cite{Gonzalez-Lopez2021,Gonzalez-Lopez2021a, Shivam2} have also shown that, in attractive systems, the elastic length scale is affected by the range of the attractive interaction.
Hence, depending on the features of the underlying energy landscape, annealing might increase or decrease the elastic length scale above which isotropic linear elasticity sets it. 


\begin{acknowledgments}
We acknowledge support from the Singapore Ministry of Education through the Academic Research Fund Tier 1 (2019-T1-001-03), Singapore and are grateful to the National Supercomputing Centre (NSCC) of Singapore for providing the computational resources.
\end{acknowledgments}


%
\end{document}